\documentclass[aip,graphicx]{revtex4-1}
\usepackage [latin1]{inputenc}
\draft 

\begin{document}

\title{Full spacetime inversion generated by electromagnetic Abelian gauge transformations} 

\author{Alcides Garat}
\affiliation{1. Former Professor at Universidad de la Rep\'{u}blica, Av. 18 de Julio 1824-1850, 11200 Montevideo, Uruguay.}
\email[]{garat.alcides@gmail.com}
\date{\today}

\begin{abstract}
\begin{center}
{\bf Abstract}
\end{center}
In a previous manuscript we addressed the possibility of generating a reflection in a region of spacetime inside a null surface under electromagnetic gauge transformations. In this manuscript we will deal with the possibility of full inversions in a region of spacetime inside a null surface under electromagnetic gauge transformations. Since new tetrads whose construction depends on electromagnetic gauge were introduced it has been proved that the physical kinematic nature of spacetime can be altered as proved previously. In this case we will see how to create a field configuration that will reverse the flow of time. We will prove that we can turn a future directed timelike vector into a past directed timelike vector by physical means. Experiments devised in order to achieve time reversal or full inversion have been never discussed before.
\end{abstract}



\maketitle 

\section{Introduction}
\label{intro}

A new kind of tetrads was introduced in a series of manuscripts \cite{A,ATGU,IWCP,LomCon,AEO,ROMP,SCR}. These tetrad vectors have remarkable properties in four-dimensional Lorentzian curved spacetimes. In Einstein-Maxwell spacetimes where we assume electromagnetic fields not to be null these properties have multiple manifestations. For example they diagonalize locally and covariantly the Einstein-Maxwell stress-energy tensor. Any stress-energy tensor, the tensor equations are general as long as the electromagnetic field is non-null. They define at each point in spacetime two orthogonal planes where every vector is an eigenvector of the stress-energy tensor. These vectors have two components in their structure. The skeletons, which are built with extremal fields. The geometrized electromagnetic fields which are second rank tensors undergo a local duality rotation through an angle local scalar that has been called the complexion and thus we obtain the extremal fields. These extremal fields are invariant under local electromagnetic gauge transformations, therefore, the skeletons are electromagnetic gauge invariants as well. Then we have the gauge vectors which are gauge by themselves and are built with gauge. The timelike and one spacelike vectors define and span the local plane one. The other two spacelike vectors define and span the orthogonal plane two. These tetrad vectors through the gauge vectors which are part of their structure, depend on gauge, in our case electromagnetic gauge. When local electromagnetic gauge transformations are carried out these vectors that span plane one undergo a hyperbolic rotation inside this local plane without leaving it plus two possible kinds of discreet transformations. Similar on the orthogonal plane two for the two spacelike vectors that undergo a spatial rotation. This means that the metric tensor is manifestly invariant under these sets of electromagnetic gauge transformations. It has been proven in detail in reference \cite{A} that the local group of the electromagnetic gauge transformations is isomorphic to both groups \cite{RGLG,RG} of tetrad transformations LB1 and LB2, independently. LB1 stands for Lorentz blade or plane one. It is a group composed by the boosts on the local plane one SO(1,1) together with two discrete transformations. One of the transformations is the full inversion or minus the identity two by two which is a Lorentz transformation. The other discrete transformation is represented by a two by two matrix and is given by $\Lambda^{o}_{\:\:o} = 0$, $\Lambda^{o}_{\:\:1} = 1$, $\Lambda^{1}_{\:\:o} = 1$, $\Lambda^{1}_{\:\:1} = 0$, that is, a reflection or flip. Since it is a flip or reflection it is not a Lorentz transformation. The group LB2 is the group of Lorentz tetrad spatial rotations on the orthogonal local plane or blade two. These results are relevant because for a long time it was thought that there was no connection between internal local gauge groups of transformations and spacetime groups of transformations. The no-go theorems from the sixties \cite{SWNG,LORNG,CMNG} established that there was no connection between the ``internal'' and the ``spacetime''. However there is a problem with the assumptions made at the outset of these theorems. We quote from reference \cite{CMNG} ``S (the scattering matrix) is said to be Lorentz-invariant if it possesses a symmetry group locally isomorphic to the Poincar\`{e} group P.\ldots A symmetry transformation is said to be an internal symmetry transformation if it commutes with P. This implies that it acts only on particle-type indices, and has no matrix elements between particles of different four-momentum or different spin. A group composed of such transformations is called an internal symmetry group''. The local electromagnetic gauge transformations are mapped into local spacetime tetrad transformations on both local orthogonal planes one and two. The Lorentz transformations on these local orthogonal planes do not commute with Lorentz transformations on other different tilted local planes. These results contradict the hypothesis made at the outset of the no-go theorems proving them incorrect. For example in reference \cite{CMNG} we can read ``Let $G$ be a connected symmetry group of the $S$ matrix, and let the following five conditions hold: 1. (Lorentz invariance) $G$ contains a subgroup locally isomorphic to $P$. 2. (Particle-finiteness) \ldots ''. If the local gauge groups of the standard model are isomorphic to the local tetrad groups of transformations LB1 or LB2 in the Abelian electromagnetic case \cite{A,ATGU,IWCP,LomCon,AEO,ROMP} or tensor products of them in the Yang-Mills general case \cite{AYM,A3,ASU3,ASUN,GIFS} then the subgroup is $G$ itself and the no-go theorems are void of any content and incorrect. These results have a profound consequence in General Relativity, Particle Physics, Gauge Theory. In addition to all these results we know from the analysis in manuscript \cite{A} that when the tetrad vectors that span the local plane one undergo a local electromagnetic gauge transformation, their norm might change in sign. There is a local scalar causality factor relating the norm of the transformed vectors and the norm of the original vectors inside the plane one such that the sign of this scalar factor can be positive or negative. These findings would be very difficult to detect without the knowledge of these new tetrads. The change in sign in the local scalar factor is associated to flips or reflections. However, we will be concerned in this manuscript with the scalar factor that does not change sign nonetheless being associated to full inversions and a jump from future directed timelike vectors into past directed timelike vectors. This will be the goal of this manuscript. We will build a model where this jump takes place from the local future light cone into the local past light cone or the intersection of the local plane one and the local future and past light cones. This is an original goal since the issue of developing experiments in order to achieve time reversal or full inversion in the plane one has been never discussed before. Throughout the paper we use the conventions of reference \cite{MW}. In particular we use a metric with sign conventions -+++. The only difference in notation with paper \cite{MW} will be that we will call our geometrized electromagnetic potential $A_{\mu}$, where $f_{\mu\nu}=A_{\nu ;\mu} - A_{\mu ;\nu}$ is the geometrized electromagnetic field $f_{\mu\nu}= (G^{1/2} / c^2) \: F_{\mu\nu}$. It is necessary to present some elements in order to address the point of this manuscript. We can proceed next to introduce the extremal field as,

\begin{equation}
\xi_{\mu\nu} = e^{-\ast \alpha} f_{\mu\nu}\ = \cos\alpha\:f_{\mu\nu} - \sin\alpha\:\ast f_{\mu\nu},\label{2dref}
\end{equation}

where $\ast f_{\mu\nu}={1 \over 2}\:\epsilon_{\mu\nu\sigma\tau}\:f^{\sigma\tau}$ is the dual tensor of $f_{\mu\nu}$. The alternating tensor $\epsilon_{\mu\nu\sigma\tau}$ is explained in section \ref{sec:appI}. Extremal fields are local electromagnetic gauge invariants as it can be noticed from equation (\ref{2dref}). Extremal fields satisfy the equation

\begin{equation}
\xi_{\mu\nu} \ast \xi^{\mu\nu}= 0\ . \label{2i0}
\end{equation}

This a condition imposed on extremal fields in order to find a local scalar called the complexion $\alpha$. The expression for the complexion, which is also a local electromagnetic gauge invariant, can be provided when imposing on the the extremal field (\ref{2dref}), the condition (\ref{2i0}), by $\tan(2\alpha) = - f_{\mu\nu}\:\ast f^{\mu\nu} / f_{\lambda\rho}\:f^{\lambda\rho}$. When we use the general identity,

\begin{eqnarray}
A_{\mu\alpha}\:B^{\nu\alpha} -
\ast B_{\mu\alpha}\: \ast A^{\nu\alpha} &=& \frac{1}{2}
\: \delta_{\mu}^{\:\:\:\nu}\: A_{\alpha\beta}\:B^{\alpha\beta}  \ ,\label{2ig}
\end{eqnarray}

which is valid for every pair of antisymmetric tensors in a four-dimensional Lorentzian spacetime \cite{MW}, and apply it to the case
$A_{\mu\alpha} = \xi_{\mu\alpha}$ and $B^{\nu\alpha} = \ast \xi^{\nu\alpha}$, it yields an equivalent condition to (\ref{2i0}),

\begin{eqnarray}
\xi_{\alpha\mu}\:\ast \xi^{\mu\nu} &=& 0\ .\label{2i2}
\end{eqnarray}

The extremal field $\xi_{\mu\nu}$ and the scalar complexion $\alpha$ have been previously defined through equations (22-25) in reference \cite{MW}. The Einstein-Maxwell stress-energy tensor according to equation (14a) in reference \cite{MW}, can be written as,

\begin{equation}
T_{\mu\nu}= f_{\mu\lambda}\:\:f_{\nu}^{\:\:\:\lambda} + \ast f_{\mu\lambda}\:\ast f_{\nu}^{\:\:\:\lambda}\ .\label{TEM}
\end{equation}

Then, the Einstein-Maxwell differential equations can be written,

\begin{eqnarray}
f^{\mu\nu}_{\:\:\:\:\:;\nu} &=& 0 \label{EM1}\\
\ast f^{\mu\nu}_{\:\:\:\:\:;\nu} &=& 0 \label{EM2}\\
R_{\mu\nu} &=& f_{\mu\lambda}\:\:f_{\nu}^{\:\:\:\lambda}
+ \ast f_{\mu\lambda}\:\ast f_{\nu}^{\:\:\:\lambda}\ , \label{EM3}
\end{eqnarray}

The local duality rotation given by equation (59) in paper \cite{MW}, the inverse of equation (\ref{2dref}) $f_{\mu\nu} = \xi_{\mu\nu} \: \cos\alpha + \ast\xi_{\mu\nu} \: \sin\alpha$, enables us to rewrite the stress-energy tensor in terms of the extremal field

\begin{eqnarray}
T_{\mu\nu}=\xi_{\mu\lambda}\:\:\xi_{\nu}^{\:\:\:\lambda} + \ast \xi_{\mu\lambda}\:\ast \xi_{\nu}^{\:\:\:\lambda} \ . \label{SETEXTREMAL}
\end{eqnarray}

It is our goal to build a tetrad in which the stress-energy tensor is diagonal. This tetrad would simplify the analysis of the geometrical properties of the electromagnetic field and also of spacetime through the metric tensor. There are four tetrad vectors that at every point in spacetime diagonalize the stress-energy tensor in Einstein-Maxwell theories for non-null electromagnetic fields,

\begin{eqnarray}
V_{(1)}^{\alpha} &=& \xi^{\alpha\lambda}\:\xi_{\rho\lambda}\:X^{\rho}
\label{V1A}\\
V_{(2)}^{\alpha} &=& \sqrt{-Q/2} \: \xi^{\alpha\lambda} \: X_{\lambda}
\label{V2A}\\
V_{(3)}^{\alpha} &=& \sqrt{-Q/2} \: \ast \xi^{\alpha\lambda} \: Y_{\lambda}
\label{V3A}\\
V_{(4)}^{\alpha} &=& \ast \xi^{\alpha\lambda}\: \ast \xi_{\rho\lambda}
\:Y^{\rho}\ ,\label{V4A}
\end{eqnarray}

where $Q=\xi_{\mu\nu}\:\xi^{\mu\nu}=-\sqrt{T_{\mu\nu}T^{\mu\nu}}$ according to equations (39) in manuscript \cite{MW}. $Q$ is assumed not to be zero, because we are dealing with non-null electromagnetic fields. Non-null we clarify means basically that $f_{\mu\nu}\:f^{\mu\nu}\neq0$ and $\ast f_{\mu\nu}\:f^{\mu\nu}\neq0$. In turn and by definitions these last equations imply that $\xi_{\mu\nu}\:\xi^{\mu\nu}\neq0$. We are free to choose the vector fields $X^{\alpha}$ and $Y^{\alpha}$, as long as the four vector fields (\ref{V1A}-\ref{V4A}) do not become trivial. Schouten defined what he called, a two-bladed structure in an Einstein-Maxwell spacetime \cite{SCH}. These local blades or planes are the planes determined by the pairs ($V_{(1)}^{\alpha}, V_{(2)}^{\alpha}$) and ($V_{(3)}^{\alpha}, V_{(4)}^{\alpha}$). The tetrad vectors have two construction components. For instance in vector $V_{(1)}^{\alpha}$ there are two main structures. First, the skeleton, in this case $\xi^{\alpha\lambda}\:\xi_{\rho\lambda}$, and second, the gauge vector $X^{\rho}$. That is, the tetrad (\ref{V1A}-\ref{V4A}) diagonalizes the stress-energy tensor for any non-trivial gauge vectors $X^{\mu}$ and $Y^{\mu}$. Two equations in the extremal field will be used in this work, in particular, to prove that tetrad (\ref{V1A}-\ref{V4A}) diagonalizes the stress-energy tensor. The first equation is given by (64) in \cite{MW}, also given in equation (\ref{2i2}). When we replace $A_{\mu\alpha} = \xi_{\mu\alpha}$ and $B^{\nu\alpha} = \xi^{\nu\alpha}$ in (\ref{2ig}), the second equation is the identity,

\begin{eqnarray}
\xi_{\mu\alpha}\:\xi^{\nu\alpha} - \ast \xi_{\mu\alpha}\: \ast \xi^{\nu\alpha} &=& \frac{1}{2} \: \delta_{\mu}^{\:\:\:\nu}\ Q \ .\label{i2}
\end{eqnarray}

When we make iterative use of (\ref{2i2}) and (\ref{i2}) we find that the vectors (\ref{V1A}-\ref{V2A}) are eigenvectors of the stress-energy tensor $T_{\mu\nu}=\xi_{\mu\lambda}\:\:\xi_{\nu}^{\:\:\:\lambda} + \ast \xi_{\mu\lambda}\:\ast \xi_{\nu}^{\:\:\:\lambda}$ with eigenvalue $\frac{Q}{2}$ while the vectors (\ref{V3A}-\ref{V4A}) are eigenvectors with eigenvalue $-\frac{Q}{2}$. Vectors (\ref{V1A}-\ref{V2A}) determine at every point in spacetime a plane one while vectors (\ref{V3A}-\ref{V4A}) determine at every point in spacetime a plane two orthogonal to plane one \cite{SCH}. The freedom we have to choose the vector fields $X^{\alpha}$ and $Y^{\alpha}$, represents available freedom that we have to choose the tetrad. If we make use of equations (\ref{2i2}) and (\ref{i2}), it is straightforward to prove that (\ref{V1A}-\ref{V4A}) is a set of orthogonal vectors. We remind ourselves that we are working with non-null electromagnetic fields. It is relevant to our work to simplify as much as we can the expression of the electromagnetic field through the use of an orthonormal tetrad, so its geometrical properties can be revealed with simplicity. As it was mentioned above we would like to show this simplification through an explicit example by making a convenient and particular choice of the vector fields $X^{\alpha}$ and $Y^{\alpha}$. In geometrodynamics, the Maxwell equations, $f^{\mu\nu}_{\:\:\:\:\:;\nu} = 0$ and $\ast f^{\mu\nu}_{\:\:\:\:\:;\nu} = 0$ have two potential vector fields associated \cite{CF}, $f_{\mu\nu} = A_{\nu ;\mu} - A_{\mu ;\nu}$ and $\ast f_{\mu\nu} = \ast A_{\nu ;\mu} - \ast A_{\mu ;\nu}$. For instance, in the Reissner-Nordstr\"{o}m geometry the only non-zero electromagnetic tensor component is $f_{tr}=A_{r;t} - A_{t;r}$ and its dual $\ast f_{\theta\phi}=\ast A_{\phi;\theta} - \ast A_{\theta;\phi}$. The symbol $``;''$ stands for covariant derivative with respect to the metric tensor $g_{\mu\nu}$ and the star in $\ast A_{\nu}$ is just a name, not the dual operator, meaning that $\ast A_{\nu ;\mu} = (\ast A_{\nu})_{;\mu}$. We understand that this notation $\ast A_{\nu}$ could be misleading but there can be no confusion since we do not mean the application of the dual Hodge operator as in the electromagnetic tensor case where $\ast f_{\mu\nu}={1 \over 2}\:\epsilon_{\mu\nu\sigma\tau}\:f^{\sigma\tau}$ is the dual tensor of $f_{\mu\nu}$, see section \ref{sec:appI}. We have been using this notation in a whole series of manuscripts \cite{A,ATGU,IWCP,LomCon,AEO,ROMP} and \cite{AYM,A3,ASU3,ASUN,GIFS} and it would be even more confusing to change it at this point. The vector fields $A^{\alpha}$ and $\ast A^{\alpha}$ represent a possible choice in Einstein-Maxwell formulations for the vectors $X^{\alpha}$ and $Y^{\alpha}$. The two vector fields do not have independence from each other, it is just a convenient choice. A further justification for the choice $X^{\alpha}=A^{\alpha}$ and $Y^{\alpha}=\ast A^{\alpha}$ could be illustrated through the Reissner-Nordstr\"{o}m geometry. In this particular geometry, $f_{tr}=\xi_{tr}$ and $\ast f_{\theta\phi}=\ast \xi_{\theta\phi}$, therefore, $A_{\theta}=0$ and $A_{\phi}=0$. Then, for the last two tetrad vectors (\ref{V3A}-\ref{V4A}), the choice $Y^{\alpha}=\ast A^{\alpha}$ becomes meaningful under the light of this particular extreme case, when basically there is no magnetic field. Then, at the points in spacetime where the set of four vectors (\ref{V1A}-\ref{V4A}) is not trivial, we can proceed to normalize,

\begin{eqnarray}
U^{\alpha} &=& \xi^{\alpha\lambda}\:\xi_{\rho\lambda}\:A^{\rho} \:
/ \: (\: \sqrt{-Q/2} \: \sqrt{A_{\mu} \ \xi^{\mu\sigma} \
\xi_{\nu\sigma} \ A^{\nu}}\:) \label{U}\\
V^{\alpha} &=& \xi^{\alpha\lambda}\:A_{\lambda} \:
/ \: (\:\sqrt{A_{\mu} \ \xi^{\mu\sigma} \
\xi_{\nu\sigma} \ A^{\nu}}\:) \label{V}\\
Z^{\alpha} &=& \ast \xi^{\alpha\lambda} \: \ast A_{\lambda} \:
/ \: (\:\sqrt{\ast A_{\mu}  \ast \xi^{\mu\sigma}
\ast \xi_{\nu\sigma}  \ast A^{\nu}}\:)
\label{Z}\\
W^{\alpha} &=& \ast \xi^{\alpha\lambda}\: \ast \xi_{\rho\lambda}
\:\ast A^{\rho} \: / \: (\:\sqrt{-Q/2} \: \sqrt{\ast A_{\mu}
\ast \xi^{\mu\sigma} \ast \xi_{\nu\sigma} \ast A^{\nu}}\:) \ .
\label{W}
\end{eqnarray}

The four vectors (\ref{U}-\ref{W}) have the following algebraic properties,

\begin{equation}
-U^{\alpha}\:U_{\alpha}=V^{\alpha}\:V_{\alpha}
=Z^{\alpha}\:Z_{\alpha}=W^{\alpha}\:W_{\alpha}=1 \ .\label{TSPAUX}
\end{equation}

Any other scalar product is zero. In order to find the expression for the electromagnetic field in terms of the tetrad (\ref{U}-\ref{W}), it is necessary to find some previous results.  The extremal field tensor and its dual can be written, see reference \cite{A},

\begin{eqnarray}
\xi_{\alpha\beta} &=& -2\:\sqrt{-Q/2}\:U_{[\alpha}\:V_{\beta]}\label{ET}\\
\ast \xi_{\alpha\beta} &=& 2\:\sqrt{-Q/2}\:Z_{[\alpha}\:W_{\beta]}\ .\label{DET}
\end{eqnarray}

Equations (\ref{ET}-\ref{DET}) are providing the necessary information to express the electromagnetic field in terms of the new tetrad,

\begin{equation}
f_{\alpha\beta} = -2\:\sqrt{-Q/2}\:\:\cos\alpha\:\:U_{[\alpha}\:V_{\beta]} +
2\:\sqrt{-Q/2}\:\:\sin\alpha\:\:Z_{[\alpha}\:W_{\beta]}\ .\label{EMT}
\end{equation}

But we can also consider Einstein-Maxwell spacetimes other than Reissner-Nordstr\"{o}m, for example perturbations to classical solutions, see reference \cite{dsmg,dsmgspringer,PIRT2019}. In that case we can make the non-trivial new choice $X^{\alpha}=A^{\alpha}$ and $Y^{\alpha}=A^{\alpha}$ and then we can explore the geometrical content of the tetrads (\ref{V1A}-\ref{V4A}) or (\ref{U}-\ref{W}). We can then explore what will be the behavior of the tetrad (\ref{V1A}-\ref{V4A}) or the normalized (\ref{U}-\ref{W}) under local electromagnetic gauge transformations. For example, from reference \cite{A} we can cite a particular boost after the gauge transformation that would look like,

\begin{eqnarray}
{\tilde{V}_{(1)}^{\alpha}
\over \sqrt{-\tilde{V}_{(1)}^{\beta}\:\tilde{V}_{(1)\beta}}}&=&
{(1+C) \over \sqrt{(1+C)^2-D^2}}
\:{V_{(1)}^{\alpha} \over \sqrt{-V_{(1)}^{\beta}\:V_{(1)\beta}}}+
{D \over \sqrt{(1+C)^2-D^2}}
\:{V_{(2)}^{\alpha} \over \sqrt{V_{(2)}^{\beta}\:V_{(2)\beta}}}\label{TN1N}\\
{\tilde{V}_{(2)}^{\alpha}
\over \sqrt{\tilde{V}_{(2)}^{\beta}\:\tilde{V}_{(2)\beta}}}&=&
{D \over \sqrt{(1+C)^2-D^2}}
\:{V_{(1)}^{\alpha} \over \sqrt{-V_{(1)}^{\beta}\:V_{(1)\beta}}} +
{(1+C) \over \sqrt{(1+C)^2-D^2}}
\:{V_{(2)}^{\alpha} \over \sqrt{V_{(2)}^{\beta}\:V_{(2)\beta}}}\ .
\label{TN2N}
\end{eqnarray}

In equations (\ref{TN1N}-\ref{TN2N}) the following notation has been used, $C = (-Q/2)\:V_{(1)\sigma}\:\Lambda^{\sigma} / (\:V_{(2)\beta}\:V_{(2)}^{\beta}\:)$, $D = (-Q/2)\:V_{(2)\sigma}\:\Lambda^{\sigma} / (\:V_{(1)\beta}\:V_{(1)}^{\beta}\:)$ and the condition $[(1+C)^2-D^2]>0$ must be satisfied. The norm of the transformed vectors
$\tilde{V}_{(1)}^{\alpha}$ and $\tilde{V}_{(2)}^{\alpha}$ is given by,

\begin{eqnarray}
\tilde{V}_{(1)}^{\alpha}\:\tilde{V}_{(1)\alpha} &=&
[(1+C)^2-D^2]\:V_{(1)}^{\alpha}\:V_{(1)\alpha}\label{FP}\\
\tilde{V}_{(2)}^{\alpha}\:\tilde{V}_{(2)\alpha} &=&
[(1+C)^2-D^2]\:V_{(2)}^{\alpha}\:V_{(2)\alpha}\ ,\label{SP}
\end{eqnarray}

where the relation $V_{(1)}^{\alpha}\:V_{(1)\alpha}=-V_{(2)}^{\alpha}\:V_{(2)\alpha}$ has been used. For the sake of simplicity the notation $\Lambda^{\alpha}$ has been used for $\Lambda^{,\alpha}$ where $\Lambda$ is the local scalar generating the local electromagnetic gauge transformation. $U^{\alpha} = {V_{(1)}^{\alpha} \over \sqrt{-V_{(1)}^{\beta}\:V_{(1)\beta}}}$ and $V^{\alpha} = {V_{(2)}^{\alpha} \over \sqrt{V_{(2)}^{\beta}\:V_{(2)\beta}}}$. For the particular case when $1+C > 0$, the transformations (\ref{TN1N}-\ref{TN2N}) reveal that an electromagnetic gauge transformation on the vector field $A^{\alpha} \rightarrow A^{\alpha} + \Lambda^{\alpha}$, that leaves unchanged the electromagnetic field $f_{\mu\nu}$, generates a boost transformation on the normalized tetrad vector fields $\left({V_{(1)}^{\alpha} \over \sqrt{-V_{(1)}^{\beta}\:V_{(1)\beta}}}, {V_{(2)}^{\alpha} \over \sqrt{V_{(2)}^{\beta}\:V_{(2)\beta}}}\right)$. In this case $\cosh(\phi) = {(1+C) \over \sqrt{(1+C)^2-D^2}}$. This was just one of the possible cases on the local plane one for proper transformations with $[(1+C)^2-D^2]>0$. Another case is for $[(1+C)^2-D^2]>0$ and $1+C < 0$ which is the composition of a boost and a full inversion.  In this case the metric tensor will satisfy the equality,

\begin{eqnarray}
g_{\mu\nu} = - U_{\mu}\:U_{\nu} + V_{\mu}\:V_{\nu} + Z_{\mu}\:Z_{\nu} + W_{\mu}\:W_{\nu} = - \widetilde{U}_{\mu}\:\widetilde{U}_{\nu} + \widetilde{V}_{\mu}\:\widetilde{V}_{\nu} + Z_{\mu}\:Z_{\nu} + W_{\mu}\:W_{\nu} \ . \label{metricproper}
\end{eqnarray}

The following notation has been used $\widetilde{U}^{\alpha}={\tilde{V}_{(1)}^{\alpha}
\over \sqrt{-\tilde{V}_{(1)}^{\beta}\:\tilde{V}_{(1)\beta}}}$ and $\widetilde{V}^{\alpha}={\tilde{V}_{(2)}^{\alpha}
\over \sqrt{\tilde{V}_{(2)}^{\beta}\:\tilde{V}_{(2)\beta}}}$. It is a direct calculation to check using the equations in reference \cite{A} or just equations (\ref{TN1N}-\ref{TN2N}) that $\widetilde{U}_{[\mu}\:\widetilde{V}_{\nu]} = U_{[\mu}\:V_{\nu]}$ regarding equation (\ref{EMT}). See reference \cite{A} section III for the detailed analysis of all possible cases. In short, the local group of electromagnetic Abelian gauge transformations is proved to be isomorphically mapped into the local group LB1 of local tetrad transformations on plane one. LB1 is made up of $SO(1,1)$ plus two different discreet transformations. One of them is the full inversion or just minus the identity two by two. The second discreet transformation is not a Lorentz transformation and is given by $\Lambda^{o}_{\:\:o} = 0$, $\Lambda^{o}_{\:\:1} = 1$, $\Lambda^{1}_{\:\:o} = 1$, $\Lambda^{1}_{\:\:1} = 0$, that is, a reflection or flip.  If the case is that $[(1+C)^2-D^2]<0$, the vectors $V_{(1)}^{\alpha}$ and $V_{(2)}^{\alpha}$ will change their timelike or spacelike character,

\begin{eqnarray}
\tilde{V}_{(1)}^{\alpha}\:\tilde{V}_{(1)\alpha} &=&
[-(1+C)^2+D^2]\:(-V_{(1)}^{\alpha}\:V_{(1)\alpha})\label{FPI}\\
(-\tilde{V}_{(2)}^{\alpha}\:\tilde{V}_{(2)\alpha}) &=&
[-(1+C)^2+D^2]\:V_{(2)}^{\alpha}\:V_{(2)\alpha}\ .\label{SPI}
\end{eqnarray}

These are improper transformations on blade one.

\begin{eqnarray}
{\tilde{V}_{(1)}^{\alpha}
\over \sqrt{\tilde{V}_{(1)}^{\beta}\:\tilde{V}_{(1)\beta}}}&=&
{(1+C) \over \sqrt{-(1+C)^2+D^2}}
\:{V_{(1)}^{\alpha} \over \sqrt{-V_{(1)}^{\beta}\:V_{(1)\beta}}}+
{D \over \sqrt{-(1+C)^2+D^2}}
\:{V_{(2)}^{\alpha} \over \sqrt{V_{(2)}^{\beta}\:V_{(2)\beta}}}\label{TN1I}\\
{\tilde{V}_{(2)}^{\alpha}
\over \sqrt{-\tilde{V}_{(2)}^{\beta}\:\tilde{V}_{(2)\beta}}}&=&
{D \over \sqrt{-(1+C)^2+D^2}}
\:{V_{(1)}^{\alpha} \over \sqrt{-V_{(1)}^{\beta}\:V_{(1)\beta}}} +
{(1+C) \over \sqrt{-(1+C)^2+D^2}}
\:{V_{(2)}^{\alpha} \over \sqrt{V_{(2)}^{\beta}\:V_{(2)\beta}}}\ .
\label{TN2I}
\end{eqnarray}

In this case the metric tensor will satisfy the equality,

\begin{eqnarray}
g_{\mu\nu} = - U_{\mu}\:U_{\nu} + V_{\mu}\:V_{\nu} + Z_{\mu}\:Z_{\nu} + W_{\mu}\:W_{\nu} = - \widetilde{V}_{\mu}\:\widetilde{V}_{\nu} + \widetilde{U}_{\mu}\:\widetilde{U}_{\nu} + Z_{\mu}\:Z_{\nu} + W_{\mu}\:W_{\nu} \ . \label{metricimproper}
\end{eqnarray}

In this case $\widetilde{V}^{\alpha} = {\tilde{V}_{(1)}^{\alpha}
\over \sqrt{-\tilde{V}_{(1)}^{\beta}\:\tilde{V}_{(1)\beta}}}$ and $\widetilde{U}^{\alpha} = {\tilde{V}_{(2)}^{\alpha}
\over \sqrt{\tilde{V}_{(2)}^{\beta}\:\tilde{V}_{(2)\beta}}}$. It is straightforward to prove using the results in reference \cite{A} or just equations (\ref{TN1I}-\ref{TN2I}) that $\widetilde{U}^{\alpha}\:\xi_{\alpha\beta}\:\widetilde{V}^{\beta} = \sqrt{-Q/2}$ implying that $\xi_{\alpha\beta} = -2\:\sqrt{-Q/2}\:\widetilde{U}_{[\alpha}\:\widetilde{V}_{\beta]}$. It is not difficult to check using the equations in reference \cite{A} or just equations (\ref{TN1I}-\ref{TN2I}) that $\widetilde{U}_{[\mu}\:\widetilde{V}_{\nu]} = U_{[\mu}\:V_{\nu]}$ regarding equation (\ref{EMT}). There is a similar analysis for the tetrad vector electromagnetic gauge transformations in the local plane two generated by ($Z^{\alpha}, W^{\alpha}$). On the local plane two the local group of electromagnetic Abelian gauge transformations is mapped isomorphically into the local group LB2 of local tetrad transformations. LB2 is $SO(2)$.  On blade or plane two, the choice $Y_{\alpha} = \ast A_{\alpha} + \ast \Lambda_{,\alpha}$ induces just local spatial rotation tetrad vector transformations,

\begin{eqnarray}
{\tilde{V}_{(3)}^{\alpha}
\over \sqrt{\tilde{V}_{(3)}^{\beta}\:\tilde{V}_{(3)\beta}}}&=&
{(1+N) \over \sqrt{(1+N)^2+M^2}}
\:{V_{(3)}^{\alpha} \over \sqrt{V_{(3)}^{\beta}\:V_{(3)\beta}}} -
{M \over \sqrt{(1+N)^2+M^2}}
\:{V_{(4)}^{\alpha} \over \sqrt{V_{(4)}^{\beta}\:V_{(4)\beta}}}\label{TN3}\\
{\tilde{V}_{(4)}^{\alpha}
\over \sqrt{\tilde{V}_{(4)}^{\beta}\:\tilde{V}_{(4)\beta}}}&=&
{M \over \sqrt{(1+N)^2+M^2}}
\:{V_{(3)}^{\alpha} \over \sqrt{V_{(3)}^{\beta}\:V_{(3)\beta}}} +
{(1+N) \over \sqrt{(1+N)^2+M^2}}
\:{V_{(4)}^{\alpha} \over \sqrt{V_{(4)}^{\beta}\:V_{(4)\beta}}}\ ,
\label{TN4}
\end{eqnarray}

where,

\begin{eqnarray}
\tilde{V}_{(3)}^{\alpha}\:\tilde{V}_{(3)\alpha} &=&
[(1+N)^2+M^2]\:V_{(3)}^{\alpha}\:V_{(3)\alpha}\label{FPS}\\
\tilde{V}_{(4)}^{\alpha}\:\tilde{V}_{(4)\alpha} &=&
[(1+N)^2+M^2]\:V_{(4)}^{\alpha}\:V_{(4)\alpha}\ ,\label{SPS}
\end{eqnarray}

and where the relation $V_{(3)}^{\alpha}\:V_{(3)\alpha}=V_{(4)}^{\alpha}\:V_{(4)\alpha}$ has been used with the following notation,

\begin{eqnarray}
M&=&(-Q/2)\:V_{(3)\sigma}\:\ast \Lambda^{\sigma} / (\:V_{(4)\beta}\:
V_{(4)}^{\beta}\:)\label{COEFFM}\\
N&=&(-Q/2)\:V_{(4)\sigma}\:\ast \Lambda^{\sigma} / (\:V_{(3)\beta}\:
V_{(3)}^{\beta}\:)\ .\label{COEFFN}
\end{eqnarray}

As long as $[(1+N)^2+M^2]>0$ the transformations (\ref{TN3}-\ref{TN4}) are telling us that an electromagnetic gauge transformation on the vector field $\ast A^{\alpha}$ that leaves invariant the dual electromagnetic field $\ast f_{\mu\nu}$, generates a spatial rotation on the normalized tetrad vector fields $\left(Z^{\alpha}={V_{(3)}^{\alpha} \over \sqrt{V_{(3)}^{\beta}\:V_{(3)\beta}}}, W^{\alpha}={V_{(4)}^{\alpha} \over \sqrt{V_{(4)}^{\beta}\:V_{(4)\beta}}}\right)$. In this case the metric tensor will satisfy the equality,

\begin{eqnarray}
g_{\mu\nu} = - U_{\mu}\:U_{\nu} + V_{\mu}\:V_{\nu} + Z_{\mu}\:Z_{\nu} + W_{\mu}\:W_{\nu} = - U_{\mu}\:U_{\nu} + V_{\mu}\:V_{\nu} + \widetilde{Z}_{\mu}\:\widetilde{Z}_{\nu} + \widetilde{W}_{\mu}\:\widetilde{W}_{\nu} \ . \label{metricproperlb2}
\end{eqnarray}

It is not difficult to verify using the equations in reference \cite{A} or just equations (\ref{TN3}-\ref{TN4}) that $\widetilde{Z}_{[\mu}\:\widetilde{W}_{\nu]} = Z_{[\mu}\:W_{\nu]}$ regarding equation (\ref{EMT}). We reiterate that local tetrad electromagnetic gauge transformations can be interpreted as new or different gauge choices $X_{\alpha} = A_{\alpha} + \Lambda_{,\alpha}$ and $Y_{\alpha} = \ast A_{\alpha} + \ast \Lambda_{,\alpha}$. All the results in section \ref{intro} are also valid for the Maxwell equations with sources $J^{\mu}$ in Minkowski spacetime were the particular tetrad construction and gauge analysis are presented in section \ref{sec:appII}. The analysis in section \ref{intro} is also valid for Einstein-Maxwell equations in curved spacetimes with sources $J^{\mu}$ were the particular tetrad construction and gauge analysis are completely analogous to the discussion presented in section \ref{sec:appII}.

\section{New physical prediction}
\label{newphysical}

If we start with a local electromagnetic gauge transformation $\Lambda$ that induces or generates a boost on the two vectors that span the local plane one as proven in detail in manuscript \cite{A} and section \ref{intro}, it has subsequently been proven that when we multiply $\Lambda$ by a suitable real constant factor $n$, the resulting local scalar $n\:\Lambda$ generates a local electromagnetic gauge transformation that inverts the future directed timelike vectors into past directed timelike vectors in a region of spacetime inside a null surface, see reference \cite{A}. There is a local scalar factor between the norm of the original vectors that span the local plane one and the transformed vectors under the corresponding local gauge transformation. This factor has been written as $[(1+C)^2-D^2]$ and it is of a kinematic nature. We call the vectors that span the plane one $V_{(1)}^{\alpha}$ and $V_{(2)}^{\alpha}$ such that both are eigenvectors of the stress-energy tensor with the same eigenvalue, see references \cite{A,ATGU,IWCP} for all the details and we also define \cite{A} the local scalars $C$ and $D$ as

\begin{eqnarray}
C&=&(-Q/2)\:V_{(1)\sigma}\:\Lambda^{\sigma} / (\:V_{(2)\beta}\:
V_{(2)}^{\beta}\:)\label{COEFFC}\\
D&=&(-Q/2)\:V_{(2)\sigma}\:\Lambda^{\sigma} / (\:V_{(1)\beta}\:
V_{(1)}^{\beta}\:)\ ,\label{COEFFD}
\end{eqnarray}

where $Q = -\sqrt{T_{\mu\nu}T^{\mu\nu}}=\xi_{\mu\nu}\:\xi^{\mu\nu}$ and $T^{\mu\nu}$ is the Einstein-Maxwell stress-energy tensor according to equations (39) in \cite{MW}. $Q$ as said before is assumed not to be zero, because we are dealing with non-null electromagnetic fields. If the case is that $[(1+C)^2-D^2]>0$, the vectors $V_{(1)}^{\alpha}$ and $V_{(2)}^{\alpha}$ will not change their timelike or spacelike character,

\begin{eqnarray}
(-\tilde{V}_{(1)}^{\alpha}\:\tilde{V}_{(1)\alpha}) &=&
[(1+C)^2-D^2]\:(-V_{(1)}^{\alpha}\:V_{(1)\alpha})\label{FPI}\\
\tilde{V}_{(2)}^{\alpha}\:\tilde{V}_{(2)\alpha} &=&
[(1+C)^2-D^2]\:V_{(2)}^{\alpha}\:V_{(2)\alpha}\ .\label{SPI}
\end{eqnarray}

These are proper transformations on blade one. By performing a local gauge transformation by a local scalar $n\:\Lambda$ in a region of spacetime inside a null surface, we are able to change timelike future oriented vectors with $1+C>0$ into timelike past oriented vectors with $1+C<0$ in the local plane one, see equations (\ref{TN1N}-\ref{TN2N}). Equivalently we can say that we manage as we will discuss in section \ref{ahatimeinv} to turn a coefficient $1+C>0$ into a coefficient $1+C<0$. The reason is that these structures arise in Einstein-Maxwell spacetimes where the notion of electromagnetic gauge is available. The fundamental point that we are making is that we can obtain this result with local Abelian gauge transformations which is a new and profound finding. By producing an electromagnetic local gauge transformation in the potentials we can change the local nature of spacetime in a region where a suitable $n\:\Lambda$ has this effect. Electromagnetic phenomena has the causality implication \cite{RW,EH,HS,CW} that enables a jump inside the region of validity from $[(1+C)^2-D^2]>0$ and $1+C>0$ into $[(1+C)^2-D^2]>0$ and $1+C<0$. This is a new and relevant development in relativity.

\section{Aharonov-Bohm setup for the full inversion}
\label{ahatimeinv}

The Aharonov-Bohm effect arises in quantum mechanics when the inclusion of a potential results in the introduction of a phase in the wave function of the electron. This phase has no consequence on the observed behavior of the electron because when a property is measured the amplitude of the wave function is involved and not its phase. Nonetheless, this phase can be detected by measuring the quantum mechanical interference between electrons that have taken two different paths from a source to a detector. If these paths travel through regions with different local values of gauge potential then a difference in phase will change the measured interference pattern. This effect was theoretically found in 1959 by Yakir Aharonov and David Bohm and confirmed by an experiment done by Robert Chambers in 1960. Chambers sent electrons on different paths that passed next to a very long solenoid. The magnetic field outside such a solenoid was negligible however the vector potential outside the solenoid was significant and local. In the end electrons taking the different paths around the solenoid acquire different phases, see references \cite{AB1,AB2,S,Y}. In this section we will be devoted to test a theoretical-experimental model for inverting the vectors in the local lightcone, see reference \cite{SCR} for the case of a spacetime reflection. To this end we will imagine to build a solenoid in the Aharonov-Bohm spirit \cite{AB1,AB2,S} with the purpose of having no electromagnetic fields outside even though the electromagnetic four-potential will not be zero, see for example reference \cite{BF}. There is a curl-free vector potential outside the solenoid with non-trivial enclosed flux. Let us also imagine that in the exterior of the solenoid \cite{NASA} we managed to create a constant magnetic field pointing in the direction of the solenoid axis. On one hand the solenoid will play the role of an Aharonov-Bohm topological element making spacetime not-simply connected. On the other hand the exterior magnetic field will make an electron move in circles of radius $R_{e}$ outside and around the solenoid such that $R_{disk} > R_{e} > R_{sol}$. $R_{sol}$ is the solenoid radius,  $R_{disk}$ will be the radius of the exterior disk where we have set up a constant possibly different magnetic field pointing parallel to the solenoid symmetry axis. The current in the solenoid coil will produce a magnetic flux that we will call $\Phi_{sol}$. Because of the non-trivial Aharonov-Bohm topology, the electron Dirac wavefunction $\Psi$ will acquire a non-trivial phase every time the electron completes a circle around the solenoid. The phase will be given by an exponential of the expression $\frac{e}{\hbar}\:\Lambda = \frac{e}{\hbar}\:\{\Phi_{sol} + \Phi_{disk}\}$. We called $e$ the charge of the electron, and $\Phi_{disk}$ the flux of magnetic field outside the solenoid in the disk region comprised between $R_{sol}$ and $R_{e}$. This way the phase acquired after one electron full rotation by the electron Dirac wave function will have two terms, one depending on time and the other on the radial coordinate, $\Lambda=\Lambda(t,r)=\Lambda_{sol}(t)+\Lambda_{disk}(r)$. We could even make if necessary for experiments $\Phi_{disk}$ the flux of magnetic field outside the solenoid in the disk region comprised between $R_{sol}$ and $R_{e}$ depend on the coordinate $z$ by making the magnetic field exterior to the solenoid and parallel to the solenoid axis such that $B_{disk}=B(z)$. The term corresponding to the exterior disk flux we will try to make of minimum influence in this problem because we are interested in the full inversion or time reversal and not on spacetime reflections or flips as in manuscript \cite{SCR}. We can do that by manipulating the exterior constant magnetic field that in this setup only satisfies the mission of making the electron move in circles around the solenoid symmetry axis. We rest our analysis on the notion that the electron Dirac wave equation will be gauge invariant, then necessarily the electromagnetic four-potential under whose presence the electron is moving will acquire a gauge term or will undergo a gauge transformation $A_{\mu} \rightarrow A_{\mu} + \Lambda_{,\mu}$. We remind ourselves that there could be different constants in the gauge transformation term, the Dirac equation minimal coupling and the exponential phase to the Dirac wavefunction according to conventions. There will be a possible non-trivial $\Lambda_{,t}$ due to the solenoid flux. We would like to have the liberty of controlling the local scalar derivative $\Lambda_{,t}$ and we can do this by changing at will the current in the solenoid coil. But most importantly we can make $\Lambda_{,t} > 0$ for pure spacetime boosts or $\Lambda_{,t} < 0$ for boosts combined with the full inversion. We can do all this increasing the current in the solenoid or decreasing the current in the solenoid, reversing the current in the solenoid, etc. This way, the $\Lambda_{sol}(t)$ term in the expression for $\Lambda=\Lambda(t,r,z)$ will control the possibility of changing the Lorentz local scalars $C$ and $D$ in the boost or in the boost combined with the full inversion. In order for these transformations to keep the timelike or spacelike character of $V_{(1)}^{\alpha}$ and $V_{(2)}^{\alpha}$ the condition $[(1+C)^2-D^2]>0$ must be satisfied. We repeat again that we are not dealing in this manuscript with reflections as in manuscript \cite{SCR} for which we would need $[(1+C)^2-D^2]<0$. If this condition $[(1+C)^2-D^2]>0$ is fulfilled, then we can normalize the transformed vectors $\tilde{V}_{(1)}^{\alpha}$ and $\tilde{V}_{(2)}^{\alpha}$ as follows,

\begin{eqnarray}
{\tilde{V}_{(1)}^{\alpha}
\over \sqrt{-\tilde{V}_{(1)}^{\beta}\:\tilde{V}_{(1)\beta}}}&=&
{(1+C) \over \sqrt{(1+C)^2-D^2}}
\:{V_{(1)}^{\alpha} \over \sqrt{-V_{(1)}^{\beta}\:V_{(1)\beta}}}+
{D \over \sqrt{(1+C)^2-D^2}}
\:{V_{(2)}^{\alpha} \over \sqrt{V_{(2)}^{\beta}\:V_{(2)\beta}}}\label{TN1}\\
{\tilde{V}_{(2)}^{\alpha}
\over \sqrt{\tilde{V}_{(2)}^{\beta}\:\tilde{V}_{(2)\beta}}}&=&
{D \over \sqrt{(1+C)^2-D^2}}
\:{V_{(1)}^{\alpha} \over \sqrt{-V_{(1)}^{\beta}\:V_{(1)\beta}}} +
{(1+C) \over \sqrt{(1+C)^2-D^2}}
\:{V_{(2)}^{\alpha} \over \sqrt{V_{(2)}^{\beta}\:V_{(2)\beta}}}\ .
\label{TN2}
\end{eqnarray}

The condition $[(1+C)^2-D^2]>0$ allows for two possible situations, $1+C > 0$ or $1+C < 0$. For the particular case when $1+C > 0$, the transformations (\ref{TN1}-\ref{TN2}) are telling us that an electromagnetic gauge transformation on the vector field $A^{\alpha}$, that leaves invariant the electromagnetic field $f_{\mu\nu}$, generates a boost transformation on the normalized tetrad vector fields $\left({V_{(1)}^{\alpha} \over \sqrt{-V_{(1)}^{\beta}\:V_{(1)\beta}}}, {V_{(2)}^{\alpha} \over \sqrt{V_{(2)}^{\beta}\:V_{(2)\beta}}}\right)$. The case $1+C < 0$, represents the composition of two transformations. An inversion of the normalized tetrad vector fields $\left({V_{(1)}^{\alpha} \over \sqrt{-V_{(1)}^{\beta}\:V_{(1)\beta}}}, {V_{(2)}^{\alpha} \over \sqrt{V_{(2)}^{\beta}\:V_{(2)\beta}}}\right)$, and a boost. After n rotations the non-trivial phase will be $n\:\Lambda$ and the non-trivial gauge transformation of the four-electromagnetic potential of the Dirac equation will be $n\:\Lambda_{,\mu}$ and the new coefficients will be $C_{new}=n\:C_{old}$ and $D_{new}=n\:D_{old}$. See equations (\ref{COEFFC}-\ref{COEFFD}) to observe that these coefficients are linear in the gradient of $\Lambda=\Lambda(t,r,z)$. We can manage at the beginning of the experiment to create a situation where $1+C_{old} > 0$ with $C_{old} < 0$ and $\Lambda_{,t} < 0$ which is just a boost. With $C_{new}=n\:C_{old}$ if $C_{old} < 0$ there will be an n such that $1+C_{new} < 0$. If the process of the electron turning around the solenoid finally makes $1+C_{new} < 0$ then we produced by physical means a full inversion combined with a boost in such a way that we reversed future timelike oriented vectors into past timelike oriented vectors. Thus producing in a region of spacetime inside a null surface a full inversion and a time inversion as we predicted. The analysis about the existence of a null surface will be carried out in detail in a following paper. We are not doing this analysis in this manuscript because it involves many non-trivial cases and it would deviate the main subject of research in this manuscript. It suffices to say that the null surface is defined by the equation $[(1+C)^2-D^2]=0$ and also see section \ref{sec:appIII} for a brief discussion.

\section{Conclusions}
\label{conclusions}

There is a new physical prediction in classical General Relativity. A new physical local quantity has been introduced. The local scalar $[(1+C)^2-D^2]$ relating the vectors that span the local plane one as in equations (\ref{FPI}-\ref{SPI}) enables a change in the very nature of spacetime. The novel result in this manuscript is related to the fact that we can control the gradient $\Lambda_{,t}$ in the general expression $\Lambda=\Lambda(t,r,z)=\Lambda_{sol}(t)+\Lambda_{disk}(r,z)$, see section \ref{ahatimeinv} and through this gradient that we can control by increasing the current in the solenoid or decreasing the current in the solenoid, reversing the current in the solenoid, etc, we can reverse the flow of time locally in a region of spacetime. These are all details associated to this unique setup and experiment. The vectors $V_{(1)}^{\alpha}$ and $V_{(2)}^{\alpha}$ might change their spacelike or timelike nature. They can also undergo a full inversion. The full inversion will transform the future directed timelike vector into a past directed timelike vector. This discrete transformation will invert the intersection of the lightcones with plane one such that at these local intersections the lightcones will be inverted. This process is enabled by several factors. First, the topology as given by Aharonov-Bohm will contribute a change in the electromagnetic gauge. This is necessary since in order to go from a setup where $1+C > 0$ into a setup where $1 + C < 0$ requires a local change in electromagnetic gauge in a region of spacetime inside a null surface. This local change in electromagnetic gauge will not alter neither the electromagnetic field nor the metric, therefore it will not change the gravitational field if spacetime is not flat, see equations (\ref{metricproper}), (\ref{metricimproper}) and (\ref{metricproperlb2}). In the Minkowski flat case the metric tensor will remain invariant as well. The Einstein-Maxwell equations on a curved four-dimensional Lorentzian spacetime are invariant under local electromagnetic gauge transformations. Second, the Dirac equation which is a quantum equation will impose invariance under local electromagnetic gauge transformations as well. It is this combination that enables the physical process of inverting the timelike vectors inside the local lightcone. We claim by using this new sector of Maxwell's ideas and the Aharonov-Bohm setup that a spinning electron around a solenoid (where in addition we have an external magnetic field to the solenoid pointing in the axis direction) in combination with the Aharonov-Bohm phase plus the ``Dirac-external electromagnetic potential-change of gauge'' can produce a change in the spacetime frame under the ``influence of the Dirac equation''. This possible change in the tetrad states could be according to the type of electromagnetic gauge transformation, any of the group tetrad transformations LB1: boosts + full inversions + spacetime reflections. The very motion of electrons around the topological element will introduce a local gauge transformation that will change the local scalar $[(1+C)^2-D^2]$ such that it might go from positive to negative in the reflection case. In the case of a full inversion combined with a boost, the local scalar $1 + C$ might go from positive to negative and $[(1+C)^2-D^2]$ still be positive and this is a unique characteristic of the results presented in this manuscript. This local scalar is not gauge invariant and that is why even though all the physical tensors and objects do not change, these local electromagnetic gauge transformations might change the nature of these local scalars. What truly changes are the local frames. They might undergo boosts, full inversions or reflections. These are very real effects that are predicted through the standard Einstein-Maxwell theories or Einstein-Maxwell equations where we can have sources or not. In fact these are also predictions of standard Maxwell electromagnetic theories with sources $J^{\mu}$ in Minkowski spacetimes because the whole formalism is also valid in flat Minkowski spacetime, for the details see section \ref{sec:appII}. The same phase effect is responsible too for the quantized-flux requirement in superconducting loops. This quantization arises because the superconducting wave function must be single valued. Around a closed loop its phase difference must be an integer multiple of $2\pi$ where the Cooper pairs have a charge $q=2e$ and therefore the flux must be a multiple of $h/2e$. The superconducting flux quantum was actually predicted before the Aharonov and Bohm ideas by F. London in 1948 using a phenomenological model \cite{FL}. Perhaps a way can be found to test the predictions of this manuscript using a superconducting flux as well. We quote from \cite{HG,WP} ``I do not doubt that classical field physics pretty directly originates from the Stoa (in ancient Greek architecture, is a covered walkway or portico, commonly for public use), in a continuous trend passing the ideas of the Renaissance and of the 17th century [\ldots]. Insofar the synthesis of quantum theory and general relativity (and, generally, field quantization) is an unsolved problem, the old (ancient) conflict between atomists and the stoics continues (p. 571)''. Up to now all these effects of a kinematic nature have been overlooked.

\section{Appendix I}
\label{sec:appI}

The Levi-Civita pseudotensor can be transformed into a tensor through
the use of factors $\sqrt{-g}$, where $g$ is the determinant of the
metric tensor. We use the notation
$e_{\alpha\beta\mu\nu}=[\alpha\beta\mu\nu]$ for the
covariant components of the Levi-Civita pseudotensor in the
Minkowskian frame given in \cite{MW},

\begin{center}
$ e_{\alpha\beta\mu\nu} = \left\{ \begin{array}{ll}
				1 \:\: \mbox{if $\alpha\beta\mu\nu$ is an even permutation of 0123}\\
				-1 \:\: \mbox{if $\alpha\beta\mu\nu$ is an odd permutation of 0123}\\
				0 \:\: \mbox{if $\alpha\beta\mu\nu$ are not all different}
				    \end{array}
			    \right. $
\end{center}

It can be noticed that the signs in $e^{\alpha\beta\mu\nu}$ are going
to be opposite to the standard notation \cite{WE}.
The reason for this is that we want to keep
the compatibility with \cite{MW} where the definition
$e_{0123}=[0123]=1$ was adopted.
With these definitions we see that in a spacetime
with a metric $g_{\alpha\beta}$,

\begin{equation}
\epsilon^{\alpha\beta\mu\nu}=
{e^{\alpha\beta\mu\nu} \over  \sqrt{-g}}=
- {[\alpha\beta\mu\nu] \over \sqrt{-g}} \ ,\label{lccon}
\end{equation}

are the components of a contravariant tensor \cite{WE,LL,MC}.
The covariant components of (\ref{lccon}) are

\begin{equation}
\epsilon_{\alpha\beta\mu\nu}= e_{\alpha\beta\mu\nu} \sqrt{-g}=
[\alpha\beta\mu\nu] \sqrt{-g} \ ,\label{lccov}
\end{equation}

where

\begin{equation}
g_{\alpha\sigma} g_{\beta\rho}
g_{\mu\kappa} g_{\nu\lambda}\:e^{\sigma\rho\kappa\lambda}= -g\:
e_{\alpha\beta\mu\nu}\ ,
\end{equation}

is satisfied.

\section{Appendix II}
\label{sec:appII}

We will study in this section how to make a suitable choice for the gauge vector $Y^{\alpha}$ for the Maxwell equations with a source $J^{\mu}$. Let us focus for practical purposes in the Coulomb problem as an example that permits a better visualization of this physical situation. The point is that in geometrodynamics, the Maxwell equations,

\begin{eqnarray}
f^{\mu\nu}_{\:\:\:\:\:;\nu} &=& J^{\mu} \label{L1}\nonumber\\
\ast f^{\mu\nu}_{\:\:\:\:\:;\nu} &=& 0 \ , \label{L2}
\end{eqnarray}

tell us about the existence of one potential $X^{\alpha}=A^{\alpha}$. Then the question arises about gauging the vectors in equations (\ref{V1A}-\ref{V4A}). The tetrad of eigenvectors to the stress-energy tensor (\ref{SETEXTREMAL}) is given by,

\begin{eqnarray}
V_{(1)}^{\alpha} &=& \xi^{\alpha\lambda}\:\xi_{\rho\lambda}\:X^{\rho}
\label{V1NONFIXEDAPP}\\
V_{(2)}^{\alpha} &=& \sqrt{-Q/2} \:\: \xi^{\alpha\lambda} \: X_{\lambda}
\label{V2NONFIXEDAPP}\\
V_{(3)}^{\alpha} &=& \sqrt{-Q/2} \:\: \ast \xi^{\alpha\lambda} \: Y_{\lambda}
\label{V3NONFIXEDAPP}\\
V_{(4)}^{\alpha} &=& \ast \xi^{\alpha\lambda}\: \ast \xi_{\rho\lambda}
\:Y^{\rho}\ .\label{V4NONFIXEDAPP}
\end{eqnarray}

When we make the gauge choice  $f_{tr} = e/r^{2}$, $A_{t} = e/r$ and $A_{r} = 0$ then $\xi_{tr}=f_{tr}$ and $\ast \xi_{\theta\phi}= \ast f_{\theta\phi}$ in a flat Minkowskian spacetime with signature $(-+++)$. We also know that the metric in spherical coordinates $(t,r,\theta,\phi)$ will be diagonal $(-1,1,r^{2},r^{2}\:\sin^{2}\theta)$. The metric determinant $g$ will satisfy $\sqrt{-g}=r^{2}\:\sin\theta$. For the alternating tensor spherical components section \ref{sec:appI} is useful when considering all these elements. In the Coulomb geometry the only non-zero tetrad vector components for the local plane one will be,

\begin{eqnarray}
V_{(1)}^{t} &=& \xi^{tr}\:\xi_{tr}\:A^{t} = \mid \xi_{tr} \mid^{2}\:A_{t} \label{V1tEX}\\
V_{(1)}^{r} &=& \xi^{rt}\:\xi_{rt}\:A^{r} = 0 \label{V1rEX}\\
V_{(2)}^{t} &=& \mid \xi_{tr} \mid \: \xi^{tr} \: A_{r} = 0 \label{V2tEX}\\
V_{(2)}^{r} &=& \mid \xi_{tr} \mid \: \xi^{rt} \: A_{t} = \mid \xi_{tr} \mid\:\xi_{tr}\:A_{t} \ .\label{V2rEX}
\end{eqnarray}

where $Q = -2\mid \xi_{tr} \mid^{2}$. In the Coulomb geometry the only non-zero tetrad vector components for the local plane two will be,

\begin{eqnarray}
V_{(3)}^{\theta} &=& \mid \xi_{tr} \mid \:\ast \xi^{\theta\phi}\:Y_{\phi} \label{V3thetaEX}\\
V_{(3)}^{\phi} &=& \mid \xi_{tr} \mid \:\ast \xi^{\phi\theta}\:Y_{\theta} = 0 \label{V3phiEX}\\
V_{(4)}^{\theta} &=& \ast \xi^{\theta\phi}\:\ast \xi_{\theta\phi}\:Y^{\theta}  = 0 \label{V4thetaEX}\\
V_{(4)}^{\phi} &=& \ast \xi^{\phi\theta}\:\ast \xi_{\phi\theta}\:Y^{\phi} \ .\label{V4phiEX}
\end{eqnarray}

We notice that using the four-dimensional Lorentz flat Minkowski metric tensor in spherical coordinates allows the introduction at every point of four orthonormal vectors. Let them be $k^{\mu}_{t}=(1,0,0,0)$, $k^{\mu}_{r}=(0,1,0,0)$, $k^{\mu}_{\theta}=(0,0,\frac{1}{r},0)$ and $k^{\mu}_{\phi}=(0,0,0,\frac{1}{r\:\sin\theta})$. Since the vector $k^{\mu}_{\phi}$ has non-trivial $\phi$ components, then we can choose the gauge vector $Y^{\alpha}=k^{\alpha}_{\phi}$. This way, the equation components (\ref{V3thetaEX}) and (\ref{V4phiEX}) will not be trivial. Despite the fact that in this Coulomb geometry we cannot choose the gauge vector $Y^{\alpha}$ to be $A^{\alpha}$ simply because the components of $V_{(3)}^{\alpha}$ and $V_{(4)}^{\alpha}$ will all be zero and nor we can choose $Y^{\alpha}$ to be $\ast A^{\alpha}$ simply because $\ast A^{\alpha}$ does not exist in the Coulomb geometry with source, we can choose it to be the vector $Y^{\alpha}=k^{\alpha}_{\phi}$. It is also possible to choose the following $Y^{\alpha}=k^{\alpha}_{\phi} + A^{\alpha}$ gauge-vector in plane two. We can then always choose another gauge by implementing $Y^{\alpha}=k^{\alpha}_{\phi} +A^{\alpha} \rightarrow Y^{\alpha}=k^{\alpha}_{\phi} +A^{\alpha} + \Lambda_{,\beta}\:g^{\alpha\beta}$. It is an electromagnetic potential gauge transformation for a valid choice of gauge-vector and we can study the tetrad eigenvector transformations in the local plane two exactly as in reference \cite{A} and section \ref{intro}. There would be no mathematical change in the analysis structure. The whole point of this section is to highlight that when we have in flat-Minkowski spacetime the Maxwell equations with sources, then we have to be careful with our choice for the gauge vector $Y^{\alpha}$. With this observation all the analysis about tetrad vector transformations in the local plane two will follow the same lines as in manuscript \cite{A} and the tetrad study originally made in section \ref{intro} for Einstein-Maxwell curved spacetimes without sources will stand. Let us remember that for Einstein-Maxwell curved spacetimes these steps were not necessary since there was a natural non-trivial choice $Y^{\alpha}=\ast A^{\alpha}$. In Einstein-Maxwell curved spacetimes with sources we would also have an analogous choice to the one given in this section depending on the case.

\section{Appendix III}
\label{sec:appIII}

In this section we will prove that the differentiation of equation (\ref{FPI}) provides the necessary information as to the nature of the null surface represented by the equation $[-(1+C)^2+D^2]=0$ and the nature of the null vectors $\tilde{V}_{(1)}^{\alpha}\:\tilde{V}_{(1)\alpha}=0$ on this surface. To this purpose let us differentiate equation (\ref{FPI}) and let us evaluate the result of the differentiation on a point $x^{\mu}$ on the null surface with respect to another point on the null surface with coordinates $x^{\mu}+dx^{\mu}$,

\begin{eqnarray}
2\:d\tilde{V}_{(1)}^{\alpha}\:\tilde{V}_{(1)\alpha} &=&
d[-(1+C)^2+D^2]\:(-V_{(1)}^{\alpha}\:V_{(1)\alpha}) + 2\:[-(1+C)^2+D^2]\:(-dV_{(1)}^{\alpha}\:V_{(1)\alpha})  \label{FPIDIFF}
\end{eqnarray}

The object $d\tilde{V}_{(1)}^{\alpha}$ represents the difference between the vector $\tilde{V}_{(1)\alpha}$ at the point $x^{\mu}+dx^{\mu}$ on the null surface and at the point $x^{\mu}$ on the null surface as well. Similar for the object $dV_{(1)}^{\alpha}$. The object $d[-(1+C)^2+D^2]$ represents the difference between the object $[-(1+C)^2+D^2]$ at the point $x^{\mu}+dx^{\mu}$ on the null surface and the point $x^{\mu}$ on the null surface as well. Then, when we evaluate both sides of (\ref{FPIDIFF}) at $x^{\mu}$,

\begin{eqnarray}
&& \left(2\:d\tilde{V}_{(1)}^{\alpha}\:\tilde{V}_{(1)\alpha}\right)|_{x^{\mu}} = \nonumber \\
&=& \left(d[-(1+C)^2+D^2]\:(-V_{(1)}^{\alpha}\:V_{(1)\alpha}) + 2\:[-(1+C)^2+D^2]\:(-dV_{(1)}^{\alpha}\:V_{(1)\alpha}\right)|_{x^{\mu}}) \ , \label{FPIDIFFATX}
\end{eqnarray}

we find as a result,

\begin{eqnarray}
2\:d\tilde{V}_{(1)}^{\alpha}\:\tilde{V}_{(1)\alpha} &=& 0 \ . \label{FPIDIFFRESULT}
\end{eqnarray}

The final result in equation (\ref{FPIDIFFRESULT}) is the consequence of $[-(1+C)^2+D^2]|_{x^{\mu}}=0$, and also $[-(1+C)^2+D^2]|_{x^{\mu}+dx^{\mu}}=0$ and hence $d[-(1+C)^2+D^2]|_{x^{\mu}}=0$. We can finally deduce that on the surface $\tilde{V}_{(1)}^{\alpha}\:\tilde{V}_{(1)\alpha}=0$ plus $d\tilde{V}_{(1)}^{\alpha}\:\tilde{V}_{(1)\alpha} = 0$ are satisfied. These last two equations mean that on one hand $\tilde{V}_{(1)}^{\alpha}$ is null on the null surface and since $d\tilde{V}_{(1)}^{\alpha}$ is parallel to the null surface because it is the difference of $\tilde{V}_{(1)\alpha}$ at the point $x^{\mu}+dx^{\mu}$ on the null surface and at the point $x^{\mu}$ also on the null surface, then we conclude that $\tilde{V}_{(1)}^{\alpha}$ is orthogonal to the null surface as expected. Similar analysis for equation (\ref{SPI}).

\section{Data availability statement}
\label{data}

The authors declare that there is no available data associated to this paper.

\section{Declaration of interests}
\label{interest}

The authors declare that they have no known competing financial interests or personal relationships that could have appeared to influence the work reported in this paper.

\end{document}